\documentclass[amsmath,amssymb,11pt,aps,a4paper]{revtex4-2}
\usepackage{amsfonts}
\usepackage[utf8]{inputenc} 
\usepackage{graphicx}
\usepackage{float}
\usepackage{natbib}
\usepackage{xcolor}
\usepackage{dcolumn}
\usepackage[english]{babel}
\usepackage{soul}
\newcommand{\be}{\begin{equation}}
	\newcommand{\ee}{\end{equation}}

\begin{document}
	
	\title{Bifurcations in the Kuramoto model with external forcing and higher-order interactions}
	
	\author{Guilherme S. Costa$^1$}
	\author{Marcel Novaes$^{2,3}$}	
	\author{Marcus A.~M.~de Aguiar$^{1,3}$}
	
	\affiliation{$^1$ ICTP South American Institute for Fundamental Research \& Instituto de F\'isica Te\'orica - UNESP, 01140-070, S\~ao Paulo, Brazil }
	\affiliation{$^2$Instituto de F\'isica, Universidade Federal de Uberl\^andia, 38408-100, Uberl\^andia, MG, Brazil}
	\affiliation{$^3$Instituto de F\'isica Gleb Wataghin, Universidade Estadual de Campinas, 13083-970, Campinas, SP, Brazil}

\begin{abstract}

Synchronization is an important phenomenon in a wide variety of systems comprising interacting oscillatory units, whether natural (like neurons, biochemical reactions, cardiac cells) or artificial (like metronomes, power grids, Josephson junctions). The Kuramoto model provides a simple description of these systems and has been useful in their mathematical exploration. Here we investigate this model combining two common \textcolor{black}{features that have been} observed in many systems: external periodic forcing and higher-order interactions among the elements. \textcolor{black}{We show that the} combination of these ingredients leads to a very rich bifurcation scenario \textcolor{black}{that produces 11 different asymptotic states of the system, with competition between forced and spontaneous synchronization. }%in the dynamics of the order parameter that describes the phase transitions. 
We found, in particular, that saddle-node, Hopf \textcolor{black}{and homoclinic} manifolds are duplicated in regions of parameter space where the unforced system displays bi-stability.
		
\end{abstract}

\maketitle

\section{Introduction}
\label{intro}

\textcolor{black}{In an influential paper published in 1967, Winfree argued that non-linear oscillators with a stable limit cycle could be characterized only by their position, or phase, along the cycle \cite{winfree1967biological}. Using approximations and numerical methods, he showed that when groups of such oscillators are coupled together, the system could undergo a transition to phase synchronization, paving the way for understanding synchrony in a variety of biological systems. Later in 1975, Kuramoto expanded on Winfree's ideas and proposed a simple model of synchronization that could be analysed in detail in the mean field limit, where the number of oscillators goes to infinity \cite{Kuramoto1975}. Kuramoto showed that synchronization would occur only if the coupling strength was larger than a minimum value, and would increase smoothly for larger coupling intensities, as in a continuous phase transition. }

%In a seminal paper published in 1967, Winfree showed that non-linear oscillators with a stable limit cycle could be characterized only by their phases along the cycle, and that interactions between such oscillators could also be described in terms of their phases \cite{winfree1967biological}. Using approximations and numerical methods, he showed that such coupled oscillators could undergo a phase transition to synchronization, paving the way for understanding synchrony in a variety of biological systems

The Kuramoto model became a paradigm in the field of synchronization and has been explored and extended in many directions
since its inception \cite{Rodrigues2016}. Examples are the introduction of frustration \cite{sakaguchi1986soluble}, networks of connections \cite{Arenas2008}, different distributions of the oscillator's natural frequencies \cite{Gomez-Gardenes2011,Ji2013,Joyce2019} and multi-dimensional models \cite{olfati2006swarms,Strogatz2019higher,barioni2021ott,Fariello2024a}. One particularly important, and largely overlooked extension, considers the dynamics of periodically forced oscillators, where mutual synchronization competes with forced entrainment \cite{Childs2008,moreira2019global,moreira2019modular}.  In many biological and artificial systems synchronization can be strongly affected by external stimuli. It is well known that photo-sensitivity to flashing lights can trigger epileptic seizures \cite{harding1994photosensitive} and that artificial pacemakers can control synchronization of heart cells \cite{Reece2012}. Information processing in the brain can also be triggered by visual, auditory or olfactory inputs \cite{Pikovsky2003}. The visual cortex, in particular, exhibits different patterns of synchronized neuronal firing when subjected to stimuli \cite{gray1994}. 

The Kuramoto model considers a set of $N$ oscillators described only by their phases and interacting with each other according to
\be \dot{\theta}_i = \omega_i + \frac{K}{N} \sum_{j=1}^N \sin{(\theta_j-\theta_i)},\ee 
where $K$ is a coupling constant. \textcolor{black}{ The natural frequencies $\omega_i$ are usually drawn from an unimodal symmetric  distribution $g(\omega)$ centered on $\omega_0$, although other arbitrary choices, such as identical oscillators, bimodal or evenly spaced frequencies \cite{ottino2016kuramoto}, have been considered.} Kuramoto showed that a phase transition exists in the large-$N$ limit, with oscillators becoming synchronized \textcolor{black}{at frequency $\omega_0$} as soon as $K$ is larger than some critical value $K_c$ that depends on $g(\omega)$. The transition may be of first or of second order, again depending on $g(\omega)$.

\textcolor{black}{The introduction of an external forcing, acting on each oscillator as $F \sin(\sigma t - \theta_i)$, corresponds to an overall influence of the environment upon the system \cite{Sakaguchi1988,Childs2008,Hindes2015,Lizarraga2020,PhysRevLett.131.030401,odor2023synchronization}.} It creates a tendency to synchronize at the frequency $\sigma$ and, therefore, competes with the internal dynamics if $\sigma\neq \omega_0$. This problem was studied by Childs and Strogatz \cite{Childs2008}, who found that the asymptotic state can undergo several phase transitions as the parameters $F$ and $\Omega=\sigma-\omega_0$ are varied. Concretely, they found explicit formulas for the boundaries between different phases in the $(F,\Omega)$ plane. These boundaries are bifurcations curves, which may be of Hopf, saddle-node, SNIPER (saddle-node with infinite period) or homoclinic type.

More recently, attention has shifted to the effects of higher order interactions \textcolor{black}{in systems of coupled oscillators \cite{tanaka2011multistable,bick2016chaos,battiston2020networks,dutta2023impact,leon2024higher}. In many complex systems interactions go beyond pairwise relations, and involve the collective action of groups of three or more agents that cannot be decomposed into a sum of pairs of interactions.} Examples can be found in neuroscience \cite{ganmor2011sparse,petri2014homological,giusti2015clique,reimann2017cliques,sizemore2018cliques}, 
ecology \cite{grilli2017higher,ghosh2024chimeric}, biology \cite{sanchez2019high} and social sciences \cite{benson2016higher,de2020social}. Studies have shown that the inclusion of higher order interactions have important consequences in the propagation of epidemics \cite{iacopini2019simplicial,jhun2019simplicial,vega2004fitness} and synchronization \cite{berec2016chimera,skardal2019abrupt,skardal2020higher,dai2021d,sarika2024,biswas2024symmetry,sayeed2024global,muolo2024phase}. 

Pairwise, or two-body, interactions corresponds to the notion of an ``interaction graph''. For higher order interactions this concept is replaced by that of a simplicial complex or hypergraph which, besides nodes and edges, also contains triangles, tetrahedra, etc. \textcolor{black}{In the Kuramoto model, 3-body interactions proportional to $\sin{(2\theta_j-\theta_k -\theta_i)}$ (asymmetric coupling) or $\sin{(\theta_j+\theta_k -2\theta_i)}$ (symmetric coupling), for example, can be added to the equation of motion of $\dot{\theta}_i$ \cite{battiston2020networks}, and justified by phase reduction methods \cite{ashwin2016hopf,leon2019phase}. Specific choices have been considered separately in order to simplify the analysis. Symmetric terms, for instance, have been studied in \cite{tanaka2011multistable,skardal2019abrupt,dai2021d,leon2024higher} whereas asymmetric interactions were considered in \cite{skardal2020higher,dutta2023impact,fariello2024third,suman2024finite}. Skardal and Arenas, in particular, have considered 3 and 4-body asymmetric interactions, with coupling constants $K_2$ and $K_3$, respectively, acting along with pairwise interactions with coupling intensity $K_1$. They showed that if the $K_2$ or $K_3$ are large enough, they can lead to bi-stability, promoted by a saddle-node bifurcation, and the appearance of a first-order phase transition \cite{skardal2020higher}. Moreover, the dynamics of the system depends on $K_2$ and $K_3$ only through their sum, $K_{23}=K_2+K_3$. %The corresponding bifurcation diagram, which contains a saddle-node line and a pitchfork line, was discussed in detail in this work.
}

Here we bring together the two features discussed above, treating a Kuramoto system \textcolor{black}{in which the oscillators are subjected to an external periodic forcing as in \cite{Childs2008} and whose internal dynamics contains higher-order} terms of the form considered in \cite{skardal2020higher}. Exploring the dynamics in full detail on the four-dimensional space of parameters $(K_1,K_{23},F,\Omega)$ is impractical, so we develop two approaches: \textcolor{black}{we investigate the effect of $K_{23}$ on the bifurcation diagram of the forced system using \cite{Childs2008} as a basis, and the effect of the forcing on the bifurcation diagram of the simplicial system, comparing the results with those in \cite{skardal2020higher}.}

\textcolor{black}{We find that $K_{23}$ produces important qualitative changes on the bifurcation diagram on the $(F,\Omega)$ plane. In particular, if $K_1<2$ and $K_{23}$ is greater than some critical value, corresponding to regions of bi-stability in the unforced system, all bifurcation curves are doubled, partitioning the parameter space into several regions of different asymptotic behaviors. We also found that any positive amount of forcing changes the nature of the pitchfork curve on the $(K_1,K_{23})$ plane, turning it into saddle-node curves. In this plane it is also possible to observe the joining of the duplicated Hopf bifurcation curves, giving rise to a Bautin critical point, where a line of saddle-node bifurcation of cycles starts.}

\textcolor{black}{The rest of the paper is organized as follows: in Section II, we describe the Kuramoto model with external forcing and higher-order interactions, deriving the dynamical equations for the order parameter using the Ott-Antonsen ansatz. In Section III, we calculate the bifurcations curves and describe the different asymptotic behaviors of system considering the simplicial parameters $K_1$ and $K_{23}$ as fixed while in Section IV we fixed the forcing parameters. Finally, in Section V we present some discussions and draw our final remarks.}

%We find that any positive amount of forcing changes the nature of the pitchfork curve on the $(K_1,K_{23})$ plane, turning it into saddle-node curves. Moreover, the bifurcation curves are changed in such a way that with increasing forcing the bi-stability region shrinks and eventually disappears for a critical value of $F$. The influence of $K_{23}$ on the bifurcation curves on the $(F,\Omega)$ plane, on the other hand, is even more dramatic. In particular, all bifurcation curves are doubled if $K_1<2$ and $K_{23}$ is larger than some critical value.

\section{The model}
\label{model}

We consider the problem of a Kuramoto model where oscillators interact in all possible pairs, triplets and quadruples, and are also subject to a periodic external force
\begin{eqnarray}
	\dot{\theta}_i &=& \omega_i + \frac{K_1}{N} \sum_{j=1}^N \sin{(\theta_j-\theta_i)} + 
	\frac{K_2}{N^2} \sum_{j,k=1}^N \sin{(2\theta_j-\theta_k -\theta_i)} \nonumber \\
	&+& \frac{K_3}{N^3} \sum_{j,k,m=1}^N \sin{(\theta_j-\theta_k +\theta_m-\theta_i)} + F \sin(\sigma t - \theta_i),
	\label{kuramoto}
\end{eqnarray}
where $K_1$ is the usual coupling constant, while $K_2$ and $K_3$ are new coupling constants that control the relative intensity of $2$-simplexes and $3$-simplexes, respectively. The natural frequencies $\omega_i$ are drawn at random from a distribution $g(\omega)$ of mean $\omega_0$ and unit variance. The particular form of higher order interactions described above is the same as in \cite{skardal2020higher} and we refer to it as \textcolor{black}
{{\it asymmetric} because $\theta_j$ and $\theta_k$ have distinct coefficients.} {\it Symmetric} higher order interactions have also been considered  \cite{skardal2019abrupt,dai2021d} but we shall not discuss them here.

We now introduce two order parameters
\begin{equation}
	z = r e^{i \psi} \equiv \frac{1}{N} \sum_{j=1}^N e^{i\theta_j},
	\label{paraord}
\end{equation}
and
\begin{equation}
	z_2 = \frac{1}{N} \sum_j e^{2 i \theta_j} \equiv r_2 e^{i \psi_2},
	\label{z2def}
\end{equation}
and change variables to $\theta_i' = \theta_i - \sigma t$, i.e. to a coordinate system that is rotating with the same frequency of the external influence. In these new coordinates, after dropping the primes, we obtain
\begin{equation}
	\dot{\theta}_i =( \omega_i -\sigma) + K_1 r \sin(\psi - \theta_i) +  K_2 r r_2 \sin{(\psi_2-\theta_i-\psi)} + K_3 r^3 \sin(\psi-\theta_i) - F \sin\theta_i.
	\label{kuramoto25}
\end{equation}

Let $f(\theta,\omega,t)$ be the density of oscillators, in the $N\to\infty$ limit, with natural frequency $\omega$ and phase $\theta$ at time $t$, so that $\int_0^{2\pi} f(\theta,\omega,t)d\theta=g(\omega)$. Using a continuity equation, the infinite system of equations of motion for the phases can be turned into an infinite system of equations of motion for the Fourier modes of $f$,
\be f=\frac{g(\omega)}{2\pi}\left(1+\sum_{n=1}^\infty f_ne^{in \theta} +c.c.\right).\ee
Upon introducing the well known Ott-Antonsen ansatz \cite{Ott2008},
\be f_n=[\alpha(\omega,t)]^n,\ee
this system collapses into a single equation for the complex parameter $\alpha$, namely
\be \dot{\alpha}=\frac{1}{2}(H^*+F)+i(\sigma-\omega)\alpha-\frac{1}{2}(H+F)\alpha^2,\ee
and
\be z=\int g(\omega)\alpha^* (\omega)d\omega,\ee
where 
\be H=K_1z+K_2z_2z^*+K_3z^2z^*.\ee

We now assume that $g(\omega)$ is a Lorentzian distribution centered on $\omega_0$ and with unit width $\Delta=1$. This allows for analytic calculations, although numerical solutions indicate that this is not essential and the main results are good approximations when other distributions with infinite support, like Gaussians, are involved. The Lorentzian assumption allows the integration to be performed by residues, in which case we have that $z_2=z^2$ and only one order parameter is required to describe the evolution of the system in the thermodynamic limit. This also means that $H=K_1z+(K_2+K_3)z|z|^2$, so only the sum of the coupling constants $K_2$ and $K_3$ is important.

This leads to a pair of coupled equations for the modulus and phase of the order parameter:
\begin{equation}\label{rdot}
	\dot{r} = -r + \frac{r}{2}(1-r^2) (K_1 + K_{23} r^2) + \frac{F}{2} (1-r^2) \cos \psi \equiv a(r,\psi)
\end{equation}
and
\begin{equation}\label{psidot}
	\dot{\psi} = -\Omega -  \frac{F}{2} \left(r+\frac{1}{r} \right) \sin \psi \equiv b(r,\psi).
\end{equation}
where $K_{23}=K_2+K_3$ and \be \Omega = \sigma-\omega_0.\ee
\textcolor{black}{Because changing $\Omega\to-\Omega$ is equivalent to changing $\psi\to -\psi$, we will restrict our attention to the case $\Omega>0$}.

The roots to the above equations, $a(r_0,\psi_0) = 0$, $b(r_0,\psi_0) = 0$, are equilibrium points of the order parameter dynamics, which may be stable or unstable. Depending on the parameters, the dynamics may be structured around cycles, with vanishing $a$ and constant $b$, and these may be attracting or repelling. Moreover, equilibrium points and cycles may be created and destroyed in bifurcations as the parameters are varied. Local information can be obtained from the Jacobian matrix,
\begin{equation}\label{jacobian}
	J = \begin{pmatrix}
		\dfrac{\partial a}{\partial r} & \dfrac{\partial a}{\partial \psi} \\[2ex]
		\dfrac{\partial b}{\partial r} & \dfrac{\partial b}{\partial \psi}
	\end{pmatrix}.
\end{equation}

A complete characterization of the dynamics in the four-dimensional space of parameters $(K_1,K_{23},F,\Omega)$ is impractical, so in the next sections we investigate the possible bifurcation scenarios under the effect of $F,\Omega$ with $K_1,K_{23}$ fixed, and under the effect of $K_1,K_{23}$ with $F,\Omega$ fixed. We restrict our attention to positive values of all parameters.

% %%%%%%%%%%%%%%%%%%%%%%%%%%%%%%%%%%%%%%%%%%%%%%%%%%%%
\section{Simplicial parameters as fixed}
\label{simp}

We consider $K_1$ and $K_{23}$ as given fixed parameters, and the forcing parameters $F$ and $\Omega$ as being variables. Our calculations are similar to the ones done in \citep{Childs2008}, which considered only the case $K_{23}=0$.

\subsection{Saddle-Node bifurcations}
\label{simpsd}

Saddle-node (SN) bifurcations are characterized by one of the eigenvalues of the Jacobian being zero at the equilibrium solutions $(r_0,\psi_0)$. The stability of the fixed points depends on the other eigenvalue of the Jacobian matrix: if positive they correspond to a saddle and an unstable node and, if negative, to a saddle and a stable node. 

The bifurcation manifold is determined by the equations
$a(r_0,\psi_0) = 0$, $b(r_0,\psi_0) = 0$, and $\det[J(r_0,\psi_0)] = 0.$ Solving these for $F$, $\Omega$ and $\psi$ leads to parametric equations in terms of the order parameter $r$:

\be F_{\rm SN}=\sqrt{\frac{2}{1-r^2}}ABr^2, \label{eqF}\ee
\be\label{eqO2} \Omega_{\rm SN}=\sqrt{\frac{(1+r^2)^{3}}{1-r^2}}2AC,\ee
and
\be \tan(\psi_{\rm SN})=\sqrt{\frac{1+r^2}{1-r^2}}\frac{C}{A},\ee
where
\be A=\sqrt{m-K_1-K_{23}r^2}, \label{eqA}\ee
\be B=\sqrt{-m^2+K_1+K_{23}(1+2r^2)},\ee
\be C=\sqrt{K_1+3K_{23}r^2-m\frac{1+r^2}{1-r^2}},\ee
and $m=2/(1-r^2)$.

\textcolor{black}{A particular case of saddle-node bifurcation occurs when the fixed points appear on a limit cycle. Right before the bifurcation the limit cycle has a very large period, that becomes infinite at the bifurcation, when one of the eigenvalues of the Jacobian is zero. The bifurcation is dubbed SNIPER, for saddle-node infinite period, or SNIC, which stands for saddle-node on an invariant cycle.}

% %%%%%%%%%%%%%%%%%%%%%%%%%%%%%%%%%%%%%%%%%%%%%%%%%%%%
% %%%%%%%%%%%%%%%%%%%%%%%%%%%%%%%%%%%%%%%%%%%%%%%%%%%%
\subsection{Hopf bifurcations}
\label{simphopf}

Hopf bifurcations requires the eigenvalues to be of the form $\pm i \omega$ at the bifurcation 
point. Therefore we impose $tr(J)=0$ and $det(J) >0$. We will deal with this last condition later. This bifurcation manifold is determined by $a(r_0,\psi_0) = 0$, $b(r_0,\psi_0) = 0$ and Tr$[J(r_0,\psi_0)] = 0.$ They can be written, after some simplification, as
\begin{equation}
	K_1 = \frac{2}{1-r^2} - \frac{F}{r} \cos\psi - K_{23} r^2,
	\label{eqH1}
\end{equation}
\begin{equation}
	\frac{F}{r} \sin\psi = - \frac{2\Omega}{1+r^2}
	\label{eqH2}
\end{equation}
and
\begin{equation}
	\frac{F}{r} \cos\psi = \frac{1}{1+3 r^2 } \left[ K_1(1-3r^2) -2 + r^2 K_{23}(3-5r^2)\right].
	\label{eqH3}
\end{equation}

Using (\ref{eqH1}) and (\ref{eqH3}) to eliminate $F \cos\phi$ we get
\begin{equation}
	K_1 = \frac{2(1+r^2)}{1- r^2 } -K_{23} r^2(2-r^2).
	\label{eqH4}
\end{equation}
This equation can be solved to get $r_{\rm Hopf}(K_1,K_{23})$. Notice that there is in general more than one solution. When $K_{23}=0$ we find, in agreement with \cite{Childs2008},
\be r_{\rm Hopf} = \sqrt{(K_1-2)/(K_1+2)}. 
\label{rhopf0}\ee

Squaring Eqs. (\ref{eqH2}) and (\ref{eqH3}) and adding them, we obtain, after some simplifications,
\begin{equation}
	F_{\rm Hopf} = r_{\rm Hopf}\left\{ \frac{4 \Omega^2}{(1+r_{\rm Hopf}^2)^2}  + \frac{r_{\rm Hopf}^4}{(1-r_{\rm Hopf}^2)^2}\left[2- K_{23}(1-r_{\rm Hopf}^2)^2\right] \right\}^{1/2}
\end{equation}
Notice that for large $\Omega$ the plot of $F(\Omega)$ can be approximated by a straight line (when $K_{23}=0$ the above expression reduces to the corresponding one in \cite{Childs2008}).

The Hopf bifurcation requires $\det(J) >0$. The point where $\det(J)=0$ is where the Hopf curve meets the saddle-node curve. This is called the Takens-Bogdanov (TB) point. The value of $\Omega$ where this occurs is given by Eq.(\ref{eqO2}) computed at $r_{\rm Hopf}(K_1,K_{23})$. 

When $K_{23} \neq 0$, Eq.(\ref{eqH4}) has to be solved numerically, considering real roots in the interval $0 < r < 1$. For $K_1 > 2$ there is always one root. For $K_1<2$ the plot of $K_{23}$ as a function of $r$ is U-shaped and there are two roots if $K_{23}$ is larger than the critical value $K_{23}^*$, where $\partial K_{23}/\partial r =0$. The roots correspond to the two Hopf curves displayed in Fig.~\ref{fig1}. At $K_{23}^*$ the two roots coalesce and disappear together with the Hopf bifurcations. For instance, when $K_1=1$ we find $K_{23}^*=5.306$. 

\subsection{Homoclinic bifurcations}

\textcolor{black}{From the TB point there also emerges the homoclinic bifurcation manifold, in which a periodic orbit collides with a saddle fixed point \cite{kuznetsov1998elements}. This curve starts tangentially to the SN curve at the TB point, where both eigenvalues of the Jacobian are zero, and also ends tangentially an another special point, the saddle-node-loop point, where the SN and SNIPER curves meet. There is no theoretical framework for analytical analysis of homoclinic curves, so we performed numerical calculations using the Julia package BifurcationKit \cite{VeltzJulia}.}

% %%%%%%%%%%%%%%%%%%%%%%%%%%%%%%%%%%%%%%%%%%%%%%%%%%%%
% %%%%%%%%%%%%%%%%%%%%%%%%%%%%%%%%%%%%%%%%%%%%%%%%%%%%
\subsection{Example: $K_1=1$ and $K_{23}=7$}
\label{simpex}

\begin{figure}[h]
	\centering
	\includegraphics[width=0.6\textwidth]{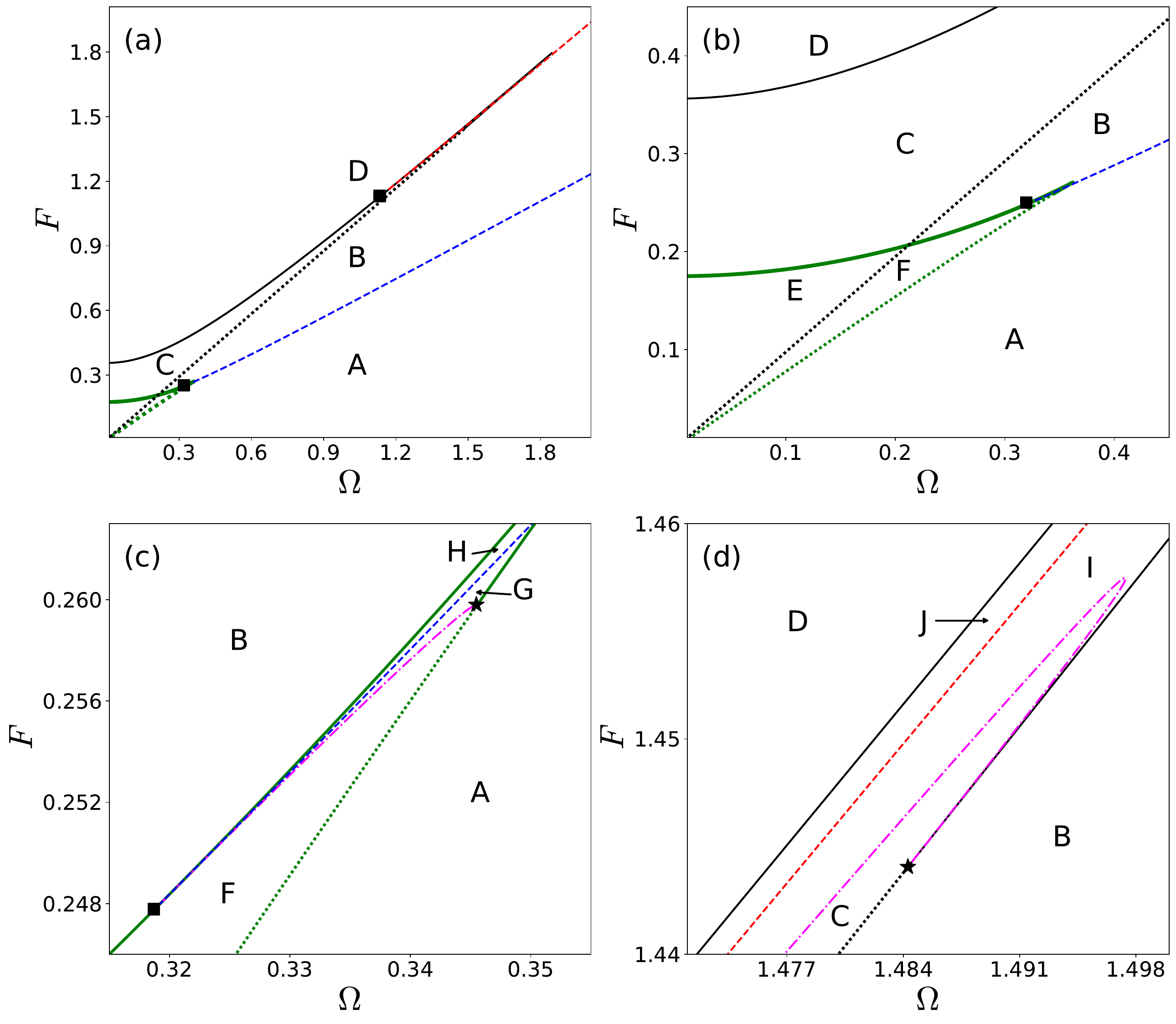}
	\caption{\textcolor{black}{Bifurcation curves in the $(F, \Omega)$ plane for $K_1 = 1$ and $K_{23} = 7$, dividing the plane into ten regions, indicated by capital letters. (a) Bifurcation diagram overview: full lines are Saddle-Node bifurcations (thin black and thick green), dotted lines are SNIPERs and dashed lines are the Hopf curves (red for super-critical and blue for sub-critical). Black squares represent Takens-Bogdanov points, where the Hopf and SN bifurcations meet. (b) Zoom of the thick green SN curve, showing in detail regions E and F. (c) Zoom on the lower Takens-Bogdanov point, with a clear view of the lower homoclinic bifurcation curve (dot-dashed pink line). (d) Zoom on the upper Takens-Bogdanov point, showing the upper homoclinic bifurcation (dot-dashed pink line). The stars mark saddle-node-loop points, where the SNIPER bifurcations turn into SN and the homoclinic bifurcation ends. 
}} 
	\label{fig1}
\end{figure}

Figure \ref{fig1} shows the bifurcation curves for $K_1=1$ and $K_{23}=7$. A dramatic difference with the case $K_{23}=0$ is that there are now two bifurcation ``tongues'' where there was previously only one: \textcolor{black}{we found two SN curves, two SNIPER curves, two Hopf curves and two homoclinic bifurcation curves (see caption for details).} When $K_{23}=0$ (only pairwise interactions) all these bifurcation manifolds exist only for $K_1>2$, as can be seen from Eq.(\ref{rhopf0}). \textcolor{black}{For $K_1<2$ there is a single stable fixed point corresponding to forced entrainment.} In the absence of forcing and higher order interactions the system recovers the Kuramoto model, which synchronizes only if $K_1 > 2$ for Lorenzian distributions with $\Delta=1$. In the next section we discuss the limit of small forcing with higher order interactions. 
%Figure \ref{fig1}(b) shows a heat map of the time averaged order parameter $r$, showing very clearly the phase transitions that take place as one crosses the bifurcation lines.

\begin{figure}[H]
	\centering
    \includegraphics[width=0.65\textwidth]{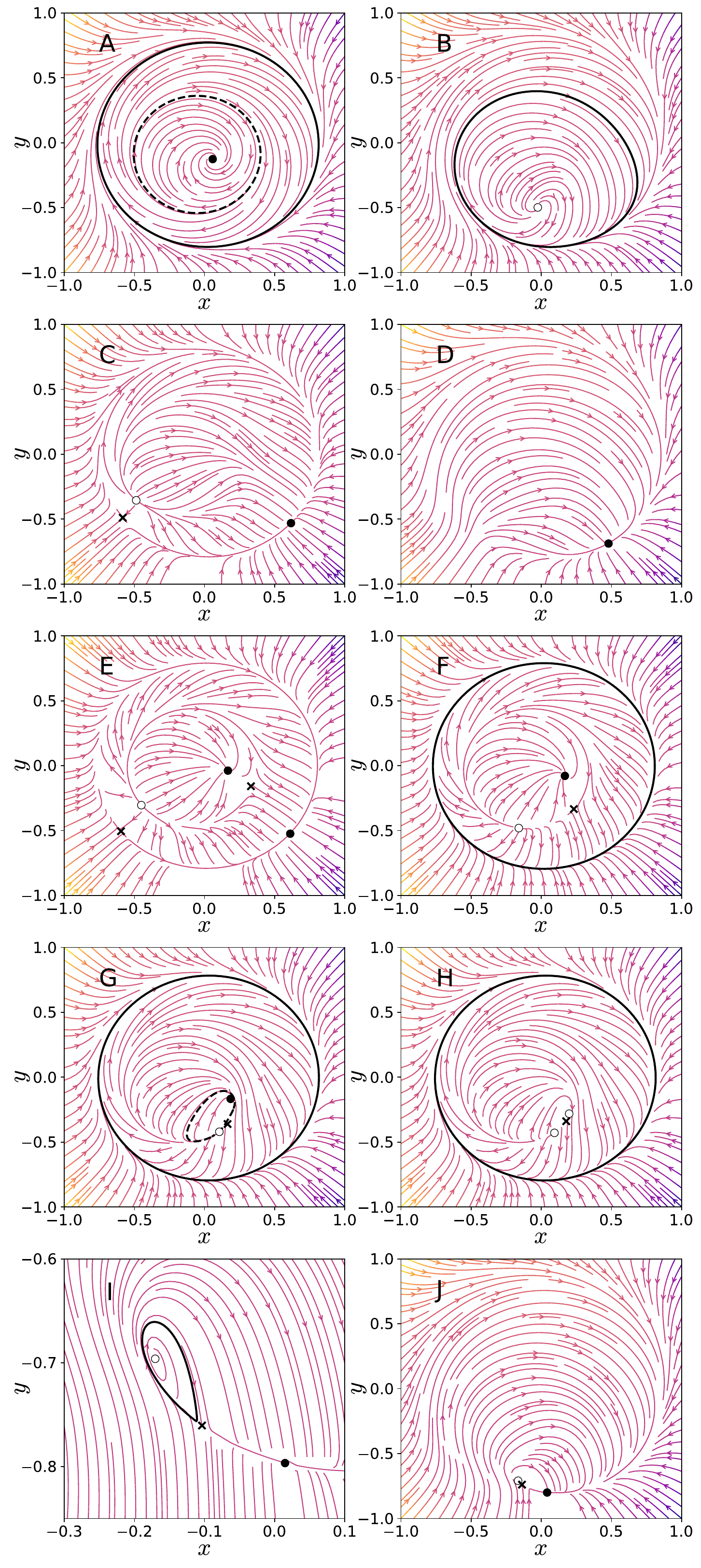}   
	\caption{\textcolor{black}{Vector fields, fixed points and periodic orbits of regions A-J shown in  Figures \ref{fig1}-(a) to (d) for $K_1=1$ and $K_{23}=7$. Full (empty) circles are stable (unstable) nodes and spirals, and crosses are saddle points. Full and dashed lines are stable and unstable orbits, respectively. \textcolor{black}{Specific parameter values for each panel are: (a) $F=0.3$, $\Omega=1.0$; (b) $F=0.3$, $\Omega=0.2$; (c) $F=0.8$, $\Omega=1.0$; (d) $F=1.2$, $\Omega=1.0$; (e) $F=0.17$, $\Omega=0.2$; (f) $F=0.15$, $\Omega=0.1$}; (g) $F=0.2612$, $\Omega=0.347$; (h) $F=0.262$, $\Omega=0.3475$; (i) $F=1.4575$, $\Omega=1.495$; (j) $F=1.4555$, $\Omega=1.4895$. }}
	\label{fig2}
\end{figure}

\textcolor{black}{The $(F,\Omega)$ plane is divided into 11 distinct regions, indicated in Figures ~\ref{fig1} and ~\ref{fig3} by capital letters. In each region the oscillators can be mutually synchronized or synchronized with the external force, exhibiting single or bi-stability, with continuous or discontinuous transitions between such states. Plots of vector fields in the $x=r \cos\psi$ and $y=r\sin\psi$ plane are shown in Figure \ref{fig2}, illustrating the behavior of the system in each region of the phase diagram and showing how limit cycles and fixed points bifurcate as $F$ and $\Omega$ change for fixed $K_1=1$ and $K_{23}=7$. Panels (a)-(j) show the vector fields in the respective regions A to J. Below we provide a brief description of the asymptotic dynamics in each of these regions. We recall that the coordinate system is rotating with the frequency of the external force, meaning that fixed points represent forced entrainment.\\}

\noindent \textcolor{black}{{\bf Region A} - The system exhibits bi-stability, with a stable fixed point close to $r=0$ (disordered motion) and a stable limit cycle with large amplitude, representing mutual entrainment of the oscillators. An unstable limit cycle separates these two stable solutions, as shown if Fig.~\ref{fig2}(a).\\}

\vspace{-0.2cm}

\noindent \textcolor{black}{{\bf Region B} - Crossing from A to B at constant $\Omega$ passes through a sub-critical Hopf bifurcation (blue dashed line) where the unstable cycle collides with the stable fixed point, that moved from the origin, and becomes unstable. The bifurcation represents a discontinuous jump from partial synchronization with the external force to highly synchronized mutual entrainment, which is now the only stable cycle, as can be seen in Fig.~\ref{fig2}(b).\\ }

\vspace{-0.2cm}

\noindent \textcolor{black}{{\bf Region C} - There is only one stable fixed point, representing forced entrainment of the oscillators. It comes about from the SNIPER bifurcation (black dotted line in Fig. \ref{fig1}) occurring on the previously stable limit cycle that existed in region B . This is illustrated in Fig.~\ref{fig2}(c), where the `ghost' of the limit cycle is still visible. \\}

\vspace{-0.5cm}

\noindent \textcolor{black}{{\bf Region D} - This is the region that dominates the plane above the diagonal, where $F$ is larger than $\Omega$. The system exhibits a single stable fixed point corresponding to forced entrainment, as shown in Fig.~\ref{fig2}(d). Moving from C to D, the unstable and saddle points shown nearby in Fig.~\ref{fig2}(c) collide in a SN bifurcation (thin black line in Fig.~\ref{fig1}). From B to D the stable cycle collapses with the unstable fixed point, in a (reversed) super-critical Hopf bifurcation. }

\vspace{-0.5cm}

\noindent \textcolor{black}{{\bf Region E} - Crossing from C to E involves the SN bifurcation represented by the thick green line in Fig.~\ref{fig1}. The bifurcation creates a pair of fixed points, a saddle and an unstable one, which can be seen close to the origin in Fig.~\ref{fig2}(e). The system exhibits bi-stability again, with a fixed point near $r=0$ (disordered motion) and a stable fixed point with large $r$ (forced entrainment). }

\vspace{-0.5cm}

\noindent \textcolor{black}{{\bf Region F} - Going from A to F or from E to F requires crossing a SNIPER bifurcation. From A to F the SNIPER happens in the unstable limit cycle and from E to F we see the collision of two points creating the stable cycle in F (reversed SNIPER). Region F is again a zone of bi-stability where disordered motion co-exists with mutual entrainment, \textcolor{black}{ as can be seen in Fig.~\ref{fig2}(f).}}

\vspace{-0.5cm}

\noindent \textcolor{black}{{\bf Region G} - Between F and G we cross a homoclinic bifurcation: starting from G, the unstable limit cycle collides with the saddle point and disappears. The system is bi-stable with a large amplitude limit cycle and a fixed point with $r \approx 0$, \textcolor{black}{see Fig.~\ref{fig2}(g)}. Region G can also be accessed from A, where a SN bifurcation creates the unstable and saddle point seen close to the unstable cycle.}

\vspace{-0.5cm}

\noindent \textcolor{black}{{\bf Region H} - Similarly to region B, the only stable state is the limit cycle, corresponding to mutual entrainment. \textcolor{black}{An extra pair of unstable and saddle fixed points can be noticed in Fig.~\ref{fig2}(h)}, due to the extra SN bifurcation at the boundary with region B. }

\vspace{-0.5cm}

\noindent \textcolor{black}{{\bf Region I} - From C to I we cross again a homoclinic bifurcation and a stable cycle appears close to the saddle point, \textcolor{black}{as seen in Fig.~\ref{fig2}(i).} The system has bi-stability with the stable cycle, representing an oscillatory forced entrainment, plus a fixed point representing forced entrainment with constant order parameter.}

\vspace{-0.5cm}

\noindent \textcolor{black}{{\bf Region J} - The subcritical Hopf bifurcation at the boundary with I makes the cycle collide with the central unstable point, leaving an unstable fixed point behind, \textcolor{black}{ see Fig.~\ref{fig2}(j).} Only the fixed point representing forced entrainment with constant order parameter remains.}

%Figure \ref{fig1}(a) also shows a grid with 8 selected points: four with $\Omega=0.2$ and four with $\Omega=1$. 

%have $\Omega=0.2$ and (e)-(h) have $\Omega=1$. Panels (d) and (h) for $F=0.02$ and $F=0.3$ respectively, show bi-stability: a vanishing order parameter, $r=0$, is a stable fixed point, but there is also a stable limit cycle with large $r$ and an unstable cycle between these two. The outer cycle represents mutual entrainment, since we are in the frame that rotates with the external force. Increasing $F$, panel (c) for $\Omega=0.2$ and $F=0.17$, shows that the unstable limit cycle has gone through a SNIPER bifurcation whereas in panel (g), for $\Omega=1$ and $F=0.8$, the unstable limit cycle has collided with the origin in a sub-critical Hopf bifurcation, where the origin becomes unstable. Only the external limit cycle is now stable.

%Increasing $F$ even more, for $\Omega=0.2$ and $F=0.3$, panel (b), two new bifurcations have occurred with respect to $F=0.17$: a SNIPER in the outer limit cycle and a saddle-node involving the origin and one of the points created at the SNIPER of the old unstable cycle. Three fixed points remain: one stable and one unstable from the SNIPER and another unstable point. For $\Omega=1$ and $F=0.99$, panel (f), we see the same structure, although the two unstable points are very close to each other and will soon disappear in another saddle-node bifurcation.

%Finally, panels (a) and (e) for large $F$ shows a single stable point corresponding to forced entrainment. 

\subsection{Evolution of Bifurcation curves}
\label{simpevol}

\begin{figure}[h]
	\centering
	\includegraphics[width=1.0\textwidth]{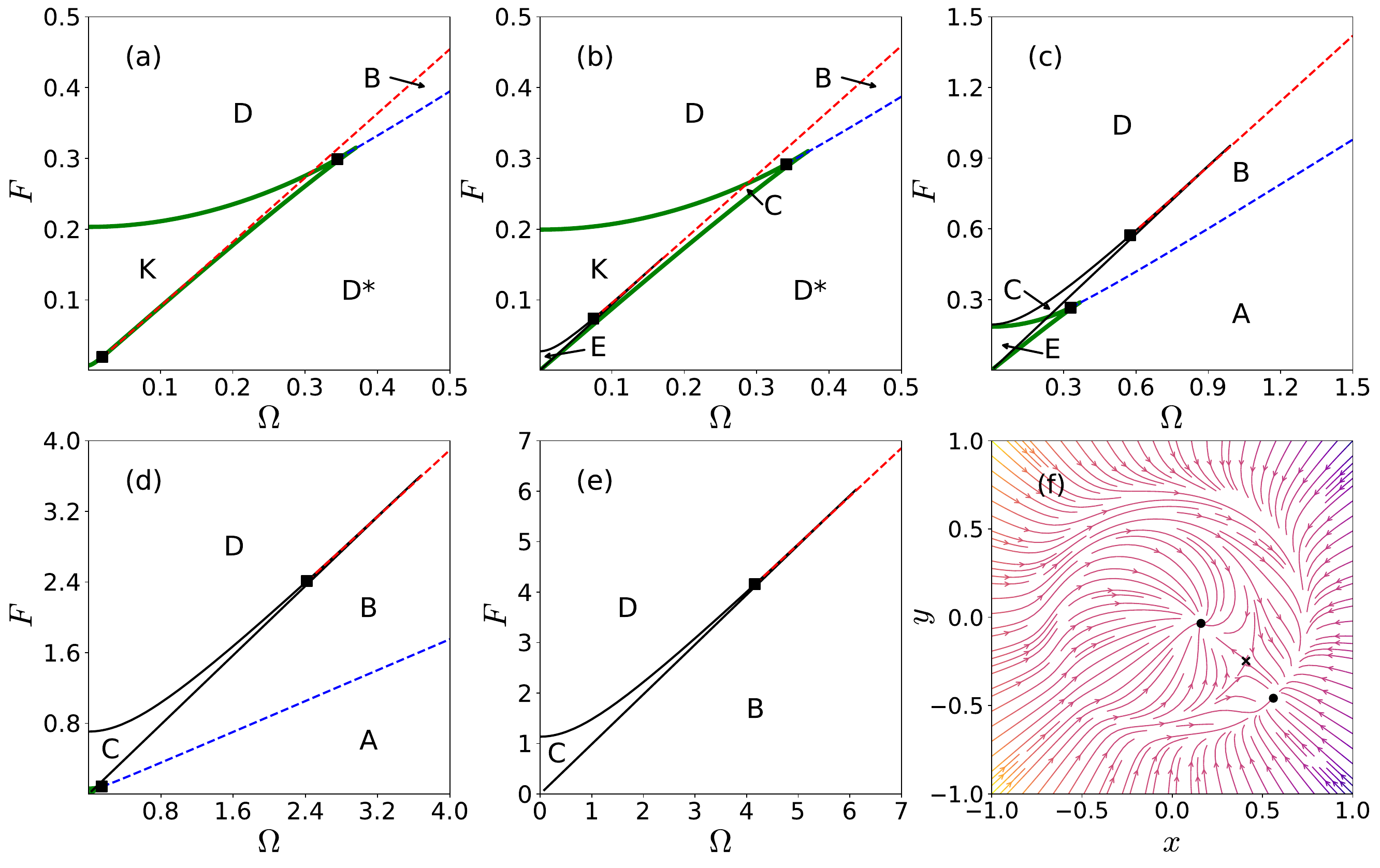}
	\caption{Saddle-node (solid) and Hopf (dashed) curves for different values of $K_1$ and $K_{23}$. In panels (a)-(c) $K_1=1$ and
	$K_{23}$ is 5.80, 5.93 and 6.5 respectively. Notice that the thick green saddle-node branch and its correspondent Hopf curve are almost independent of $K_{23}$, whereas the other ones grow as $K_{23}$ increases. In (d)-(e) $K_{23}=7$ and $K_1$ is 1.5 and 2.1 respectively. The thick green branch disappears at $K_1=2$. Black squares show Takens-Bogdanov points. Panel (f) shows the vector field and fixed points found in Region $K$ ($F=0.15$, $\Omega = 0.1$, $K_1=1$, $K_{23}=5.93$).} 
	\label{fig3}
\end{figure}

\textcolor{black}{As $K_1$ and $K_{23}$ varies, the projection of the bifurcation manifolds on the $(F,\Omega)$ plane changes. Figure \ref{fig3} illustrates how SN and Hopf bifurcation curves change for a few selected values of $K_1$ and $K_{23}$. For the sake of simplicity we shall refer to the two SN branches as the {\it green} (thick green line) and {\it black} (thin black line) branches, referring to the colors displayed in Fig.~\ref{fig1}. Also, for the sake of clarity and lack of resolution in the figures, we do not display the homoclinic bifurcation curves.} For $K_1=1$, panels (a)-(c), the green branch of the SN manifold is almost independent of $K_{23}$, whereas the black branch, created at $K_{23} \approx 5.83$ (see below), grows as $K_{23}$ increases. For $K_{23} =5.8$, panel (a), the black branch has collapsed into the green branch, but its Takens-Bogdanov point remains with its associated Hopf curve. Panels (d)-(e) have $K_{23}=7$ and increasing values of $K_1$. The green branch now shrinks and disappears at $K_1=2$. For $K_1>2$ only one branch and its attached Hopf curve survive, similar to the case $K_{23}=0$.
% %%%%%%%%%%%%%%%%%%%%%%%%%%%%%%%%%%%%%%%%%%%%%%%%%%%%
% %%%%%%%%%%%%%%%%%%%%%%%%%%%%%%%%%%%%%%%%%%%%%%%%%%%%
 \textcolor{black}{Two new regions were identified with respect to the classification in Figs.~\ref{fig1} and \ref{fig2}. Region $D^*$ differs from $D$ by having the only stable fixed point close to the origin, indicating disordered motion. Figure 3(f) shows the vector field in the novel region K that appears in panels 3(a) and (b):\\}

\noindent \textcolor{black}{{\bf Region K} - Bi-stability between disordered motion and forced entrainment, similar to region $E$, but without the unstable and saddle fixed points observed at the left of Fig.~\ref{fig2}(e), that collide on the SN bifurcation.\\}

More information about the critical values described above can be obtained by setting $F=\Omega=0$ in Eqs. (\ref{eqF}) and (\ref{eqO2}). This amounts to set $A=0$, Eq.(\ref{eqA}). Solving for $K_{23}$ we find $K_{23} = (m-K_1)/r^2$, where $m=2/(1-r^2)$. As a function of $r$, $K_{23}$ is U-shaped in the interval $(0,1)$. Imposing $\partial K_{23}/\partial r = 0$ at $r=r_c$ we obtain
\be r_c = 1-\frac{2}{K_1}+ \frac{2}{K_1}\sqrt{1-\frac{K_1}{2}}. \ee
For $K_1=1$ we find $r_c=\sqrt{2}-1$ and $K_{23}(r_c) \approx 5.83$. Below this point the two legs of the black branch collapse into the lower end of the green branch (see Fig.~\ref{fig3}(a)). Notice that $K_{23}(r_c)$ is larger than $K_{23}^* \approx 5.306$, which means that two Hopf bifurcation curves exist attached to a single SN tongue. This phenomena will occur for $K_1=1$ as long as $K_{23}$ is in the [5.306,  5.83] interval. Decreasing $K_{23}$ even more would show the two Takens-Bogdanov points coalescing at $K_{23}^*$ and the Hopf curves disappearing. 

Finally, we see that $K_1=2$ is also a critical point, where $r_c=0$ and the green branch disappears. The two branches coexist only for $K_1 < 2$ and $K_{23} > K_{23}(r_c)$.

\section{Forcing parameters as fixed}
\label{force}

Now we consider $F$ and $\Omega$ as given fixed parameters and vary the coupling parameters $K_1$ and $K_{23}$. Our calculations are similar to the ones done in \citep{skardal2020higher}, which considered only the case $F=\Omega=0$. 

The authors of \citep{skardal2020higher} found a pitchfork bifurcation for $K_1=2$, independent of $K_{23}$, and a saddle-node bifurcation along a certain curve in the $(K_1, K_{23})$ plane. It turns out that the situation in the presence of forcing is quite different: for any finite value of the forcing parameters, the pitchfork bifurcation is replaced by a saddle-node curve, together with Hopf and homoclinic bifurcation lines; moreover, this structure may appear duplicated.

In this case, by solving the saddle-node conditions $a(r_0,\psi_0) = 0$, $b(r_0,\psi_0) = 0$ and $\det[J(r_0,\psi_0)] = 0$ for $K_1$, $K_{23}$ and $\phi$, we get the parametric solutions
\be K_{1,\rm SN} =-\frac{3F^2(1+r^2)}{2Dr}+\frac{4\Omega^2 r(1+2r^2)}{D(1+r^2)^2}+\frac{2(1-2r^2)}{(1-r^2)^2},\ee
\be K_{23,\rm SN}=\frac{F^2(1+r^2)}{2Dr^3}-\frac{4\Omega^2 r}{D(1+r^2)^2}+\frac{2}{(1-r^2)^2},\ee
\be \tan(\psi_{\rm SN})=\frac{2\Omega r}{D} \quad (-\pi<\psi_{\rm SN}<0),\ee
where
\be D=\pm\sqrt{F^2(1+r^2)^2-4\Omega^2 r^2},\ee
with the plus and minus solutions generating two different branches of the bifurcation curve.

For the Hopf bifurcation, on the other hand, it is the trace of the Jacobian that vanishes, instead of the determinant. This leads to   
\be\label{KHopf} K_{1,\rm Hopf}=\frac{2(1-2r^2)}{(1-r^2)^2}+\frac{D(2-r^2)}{r(1-r^4)},\ee
\be \label{HHopf}K_{23,\rm Hopf}=\frac{2}{(1-r^2)^2}-\frac{D}{r^3(1-r^4)},\ee
\be \tan(\psi_{\rm Hopf})=\frac{2\Omega r}{D} \quad (-\pi<\psi_{\rm Hopf}<0),\ee
where again we have two branches.

The determinant of the Jacobian must be positive at the Hopf manifold, which happens to one side of a Takens-Bogdanov point. To find such a point, we must solve $\dot{r}=\dot{\psi}={\rm Tr}(J)=\det(J)=0$, simultaneously.
We find that the easiest way to do this is to treat $\cos\psi$ and $\sin\psi$ as formally independent, solve for $K_1,K_{23},\cos\psi,\sin\psi$, and then impose $\sin^2\psi+\cos^2\psi=1$, as done in \cite{Childs2008}. For given values of $(F,\Omega)$, the value of $r$ which corresponds to the Takens-Bogdanov point satisfies
\be F^2(1+r_{\rm TB}^2)^4=8\Omega^2 r_{\rm TB}^2(1+r_{\rm TB}^4),\ee
and its coordinates in the $(K_1,K_{23})$ plane are given by (\ref{KHopf}) and (\ref{HHopf}) computed at  $r_{\rm TB}$. 

\begin{figure}[h]
	\centering
	\includegraphics[width=0.95\textwidth]{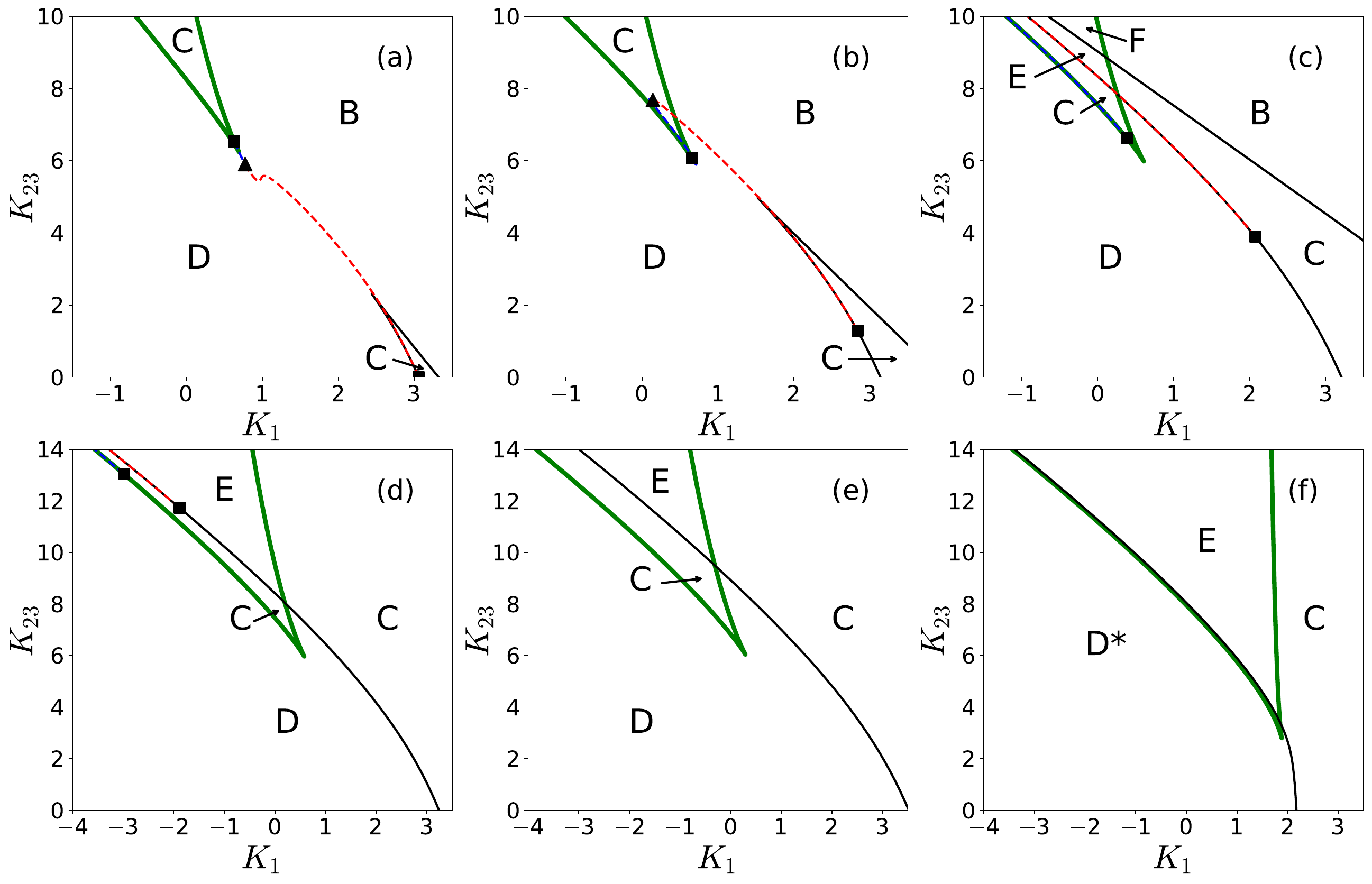}
	\caption{Saddle-node and Hopf curves for different values of $F$ and $\Omega$. In panels (a)-(e) $\Omega=0.5$ and
	$F$ is 0.45, 0.47, 0.49, 0.499 and 0.59, respectively. In panel (f), $\Omega = 0.01$ and $F = 0.02$.  Black squares show Takens-Bogdanov points \textcolor{black}{and triangles are Bautin points, where sub and super-critical Hopf bifurcation curves meet (see Fig.~\ref{fig5}(a)). Regions are labeled according to Figs.~\ref{fig1}, \ref{fig2} and \ref{fig3}.} }
	\label{fig4}
\end{figure}

\begin{figure}[h]
	\centering
	\includegraphics[width=0.8\textwidth]{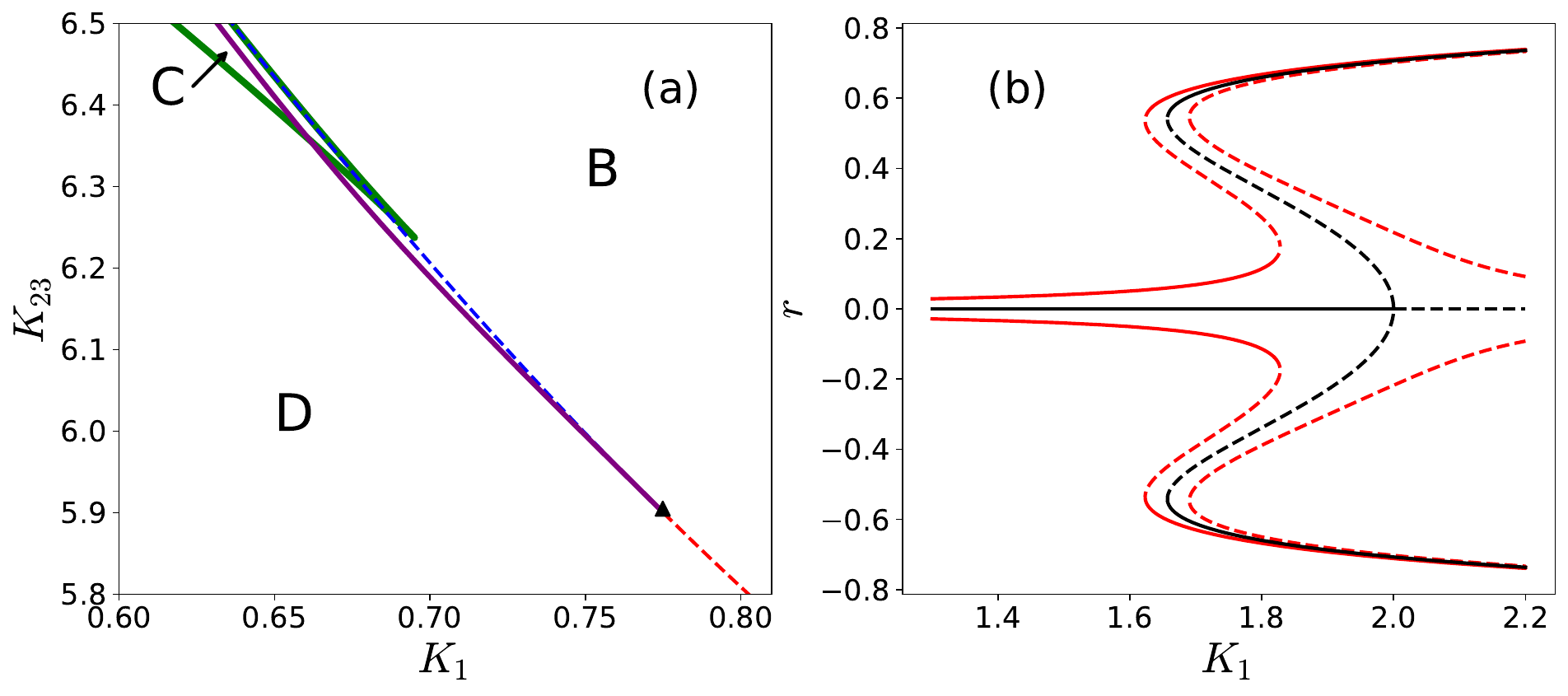}
	\caption{(a) Zoom of Fig.~\ref{fig4}(a) showing the joining of sub-critical (blue) and super-critical (red) Hopf bifurcation manifolds at the Bautin point (triangle). A line of saddle-node bifurcation of cycles (purple) starts from the same point. (b) Equilibrium values of the order parameter $r$ as function of $K_1$ for $K_{23}=4$. Solid lines for stable equilibria, dashed lines for unstable ones. Black lines display a pitchfork bifurcation for $F=0$. For $F=0.02$ and $\Omega=0.01$, red lines, the pitchfork unfolds into saddle-node bifurcations. Values of $r<0$ are shown for the sake of visualization. }
	\label{fig5}
\end{figure}

Since $r^2(1+r^4)/(1+r^2)^4\le 1/8$ for $0\le r\le 1$, the above equation only has relevant solutions if $F\le \Omega$. In general, each Hopf branch has its TB point, and the values of $\phi_{\rm TB}$ at these different points are related by $\psi_{\rm TB}^{(2)}=\pi-\psi_{\rm TB}^{(1)}$.
%\textcolor{black}{Comentar sobre a bifurcação GH e a curva T aqui \cite{kuznetsov1998elements} ja coloquei a ref do livro.}

\textcolor{black}{Figure 4 shows the bifurcation curves in the $(K_1,K_{23})$ plane, for a few values of $F$ and $\Omega$. In panels (a)-(e), $\Omega=0.5$ and $F$ takes the values $0.45$, $0.47$, $0.49$, $0.499$ and $0.59$, respectively.} \textcolor{black}{Panel (a) shows the two saddle-node branches (thin black and thick green solid lines) connected by the red (super-critical) and blue (critical) Hopf branches. The point where the Hopf curves join is $K_{23}^*$, as discussed in section \ref{simphopf}, and characterize a Generalized Hopf, or Bautin, bifurcation, where a line of SN of limit cycles also issues \cite{kuznetsov1998elements}, as shown in Fig.~\ref{fig5}(a)}. 

\textcolor{black}{As $F$ grows, the lower SN branch grows and crosses the upper one, which slowly recedes to negative values of $K_1$. When $F\to \Omega$, the TB points slide off to infinity. In panel 4(f) we show that for small values of $F$ and $\Omega$, the bifurcation diagram from \cite{skardal2020higher} is recovered. However, for any $F>0$ the pitchfork bifurcation unfolds into SN curves, as shown if Fig.~\ref{fig5}(b), where we plot all equilibrium values of $r$ that exist for $K_{23}=4$ as functions of $K_1$. As a consequence, $r=0$ is no longer an equilibrium.}

%\begin{figure}[h]
	%\centering
%	\includegraphics[width=0.4\textwidth]{F045-Om05.pdf}
	%\caption{Heat map of the time averaged order parameter $\langle r \rangle$ showing the phase transitions promoted by the bifurcation lines, for $\Omega=0.5$ and $F=0.45$, as in fig. 4a.} 
	%\label{fig5}
%\end{figure}

%Figure 5 shows a heat map of the time averaged order parameter $\langle r\rangle$, along with the bifurcation lines. The transitions in this case are not as abrupt as they appear in Figure 1-(b), with the exception of one of them, in the upper branch of the saddle-node manifold. The Hopf bifurcations, on the other hand, are not very noticeable, because in this case what happens is that a stable fixed point with $\langle r\rangle > 0$ becomes unstable and a periodic orbit is created which has approximately the same $\langle r\rangle$.

\section{Conclusions}
\label{conc}

\textcolor{black}{Coupled oscillators are often subjected to external forcing, introducing a competition between spontaneous and forced synchronization. Examples include artificial pacemaker devices implanted on the heart \cite{Reece2012}, responses of the visual cortex to stimuli \cite{gray1994} and epileptic seizures induced by flashing lights \cite{harding1994photosensitive}. The importance of understanding forced systems of oscillators, therefore, goes much beyond pure mathematical interest. Child and Strogatz \cite{Childs2008} used the Kuramoto model to study the effects of a periodic force. They showed that the asymptotic states depended on the force intensity $F$ and on the mismatch between the external frequency and average natural frequency of the oscillators, $\Omega$. For fixed $\Omega$ and small $F$, spontaneous synchronization is observed, whereas for sufficiently large $F$ the oscillators synchronize with the external force. More importantly, the transition between the two regimes displayed a rich set of bifurcations with periodic cycles and equilibrium points that depended very sensitively on the values of $F$, $\Omega$ and the coupling parameter $K$. }

In this work we extended the bifurcation analysis of the forced Kuramoto model to include higher order interactions between the oscillators. The effects of higher order terms in states of the Kuramoto model was discussed by Skardal and Arenas, who showed that the contribution of third and forth order interactions could be put together in terms of a single parameter, $K_{23}$, but did not considered external forces \cite{skardal2020higher}. Here we described in detail how these interactions impact the bifurcation diagram obtained originally in \cite{Childs2008} in the $(F,\Omega)$ plane and also how the intensity and frequency of the external force change the diagram described in \cite{skardal2020higher} in the $(K_1,K_{23})$ plane. 

\textcolor{black}{We have shown that two branches of the saddle-node (SN) bifurcation manifold can co-exist for some values of the parameters, reflecting the bi-stability found in \cite{skardal2019abrupt} for the unforced case. Lines of Hopf and homoclinic bifurcations issue from Takens-Bogdanov points, dividing the space of parameters in many regions of quite distinct dynamical character, as illustrated in Figs.~\ref{fig1}-\ref{fig4}. The two branches of Hopf bifurcations meet at the so called generalized Hopf or Bautin bifurcation, point where a line of saddle-node of cycles also starts. The combined effects of higher order interactions and external periodic force increases dramatically the complexity of the bifurcation diagrams, that is divided into 11 different topological regions, making the determination of the asymptotic state of the system very sensitive to parameter values.}

\textcolor{black}{Finally, we note that higher order interactions, as derived from phase reduction methods \cite{ashwin2016hopf,leon2019phase}, can have different functional forms, and a study of their combined effects has not yet been investigated in detail. Instead, the effect of specific terms have been studied separately, revealing distinct patterns and states. Asymmetric interaction terms are amenable to exact treatment under the Ott-Antonsen ansatz, and have been analysed in refs. \cite{skardal2019abrupt,dutta2023impact} and here. The Kuramoto model with symmetric higher order terms, on the other hand, exhibits multi-stability  \cite{tanaka2011multistable,skardal2019abrupt,dai2021d,skardal2020higher} and anomalous transitions to synchrony \cite{leon2024higher}. In this case, the effects of external forcing might lead to a proliferation of branches in the bifurcation manifolds. We leave the  exploration of these cases for future work. }

%It would be possible to consider interactions of even higher orders, involving more than four oscillators. This could lead to increasing non-linearity in the equations of motion of the order parameter and a proliferation of branches in the bifurcation manifolds and the phenomenon of multi-stability \cite{dai2021d}. However, high order interactions can be defined in many different ways, and most of them are not amenable to exact treatment under the Ott-Antonsen ansatz. We leave such an exploration for future work.

%%%%%%%%%%%%%%%%%%%%%%%%%%%%%%%%%%%%%%%%%%%%%%%%%%%%
%%%%%%%%%%%%%%%%%%%%%%%%%%%%%%%%%%%%%%%%%%%%%%%%%%%%
\begin{acknowledgments}
	We would like to thank the anonymous referees for their valuable remarks. This work was partly supported by FAPESP, grants 2021/14335-0 (MAMA), 2023/03917-4 (GSC) and 2023/15644-2 (MN), and also by CNPq, grants 301082/2019-7 (MAMA) and 304986/2022-4 (MN).
\end{acknowledgments}

%%%%%%%%%%%%%%%%%%%%%%%%%%%%%%%%%%%%%%%%%%%%%%%%%%%%%

\clearpage
\enlargethispage{1\baselineskip} 
\newpage
\bibliographystyle{ieeetr}

\begin{thebibliography}{10}
	
	\bibitem{winfree1967biological}
	A.~T. Winfree, ``Biological rhythms and the behavior of populations of coupled
	oscillators,'' {\em Journal of theoretical biology}, vol.~16, no.~1,
	pp.~15--42, 1967.
	
	\bibitem{Kuramoto1975}
	Y.~Kuramoto, ``{Self-entrainment of a population of coupled non-linear
		oscillators},'' in {\em International Symposium on Mathematical Problems in
		Theoretical Physics}, pp.~420--422, Berlin/Heidelberg: Springer-Verlag, 1975.
	
	\bibitem{Rodrigues2016}
	F.~A. Rodrigues, T.~K. D.~M. Peron, P.~Ji, and J.~Kurths, ``{The Kuramoto model
		in complex networks},'' {\em Physics Reports}, vol.~610, pp.~1--98, 2016.
	
	\bibitem{sakaguchi1986soluble}
	H.~Sakaguchi and Y.~Kuramoto, ``A soluble active rotater model showing phase
	transitions via mutual entertainment,'' {\em Progress of Theoretical
		Physics}, vol.~76, no.~3, pp.~576--581, 1986.
	
	\bibitem{Arenas2008}
	A.~Arenas, A.~Diaz-Guilera, J.~Kurths, Y.~Moreno, and C.~Zhou,
	``{Synchronization in complex networks},'' {\em Physics Reports}, vol.~469,
	pp.~93--153, 2008.
	
	\bibitem{Gomez-Gardenes2011}
	J.~Gomez-Gardenes, S.~Gomez, A.~Arenas, and Y.~Moreno, ``{Explosive
		synchronization transitions in scale-free networks},'' {\em Physical Review
		Letters}, vol.~106, no.~12, pp.~1--4, 2011.
	
	\bibitem{Ji2013}
	P.~Ji, T.~K.~D. Peron, P.~J. Menck, F.~A. Rodrigues, and J.~Kurths, ``{Cluster
		explosive synchronization in complex networks},'' {\em Physical Review
		Letters}, vol.~110, no.~21, pp.~1--5, 2013.
	
	\bibitem{Joyce2019}
	J.~S. Climaco and A.~Saa, ``Optimal global synchronization of partially forced
	kuramoto oscillators,'' {\em Chaos: An Interdisciplinary Journal of Nonlinear
		Science}, vol.~29, no.~7, p.~073115, 2019.
	
	\bibitem{olfati2006swarms}
	R.~Olfati-Saber, ``Swarms on sphere: A programmable swarm with synchronous
	behaviors like oscillator networks,'' in {\em Proceedings of the 45th IEEE
		Conference on Decision and Control}, pp.~5060--5066, IEEE, 2006.
	
	\bibitem{Strogatz2019higher}
	M.~Lipton, R.~Mirollo, and S.~H. Strogatz, ``On higher dimensional generalized
	kuramoto oscillator systems,'' {\em arXiv preprint arXiv:1907.07150}, 2019.
	
	\bibitem{barioni2021ott}
	A.~E.~D. Barioni and M.~A. de~Aguiar, ``Ott--antonsen ansatz for the
	d-dimensional kuramoto model: A constructive approach,'' {\em Chaos: An
		Interdisciplinary Journal of Nonlinear Science}, vol.~31, no.~11, p.~113141,
	2021.
	
	\bibitem{Fariello2024a}
	R.~Fariello and M.~A. de~Aguiar, ``Exploring the phase diagrams of
	multidimensional kuramoto models,'' {\em Chaos, Solitons \& Fractals},
	vol.~179, p.~114431, 2024.
	
	\bibitem{Childs2008}
	L.~M. Childs and S.~H. Strogatz, ``{Stability diagram for the forced Kuramoto
		model},'' {\em Chaos}, vol.~18, no.~4, pp.~1--9, 2008.
	
	\bibitem{moreira2019global}
	C.~A. Moreira and M.~A. de~Aguiar, ``Global synchronization of partially forced
	kuramoto oscillators on networks,'' {\em Physica A: Statistical Mechanics and
		its Applications}, vol.~514, pp.~487--496, 2019.
	
	\bibitem{moreira2019modular}
	C.~A. Moreira and M.~A. de~Aguiar, ``Modular structure in c. elegans neural
	network and its response to external localized stimuli,'' {\em Physica A:
		Statistical Mechanics and its Applications}, vol.~533, p.~122051, 2019.
	
	\bibitem{harding1994photosensitive}
	G.~F. Harding and P.~M. Jeavons, {\em Photosensitive epilepsy}.
	\newblock No.~133, Cambridge University Press, 1994.
	
	\bibitem{Reece2012}
	J.~B. Reece, {\em {Campbell biology : concepts {\&} connections}}.
	\newblock San Francisco, CA.: Benjamin Cummings, 2012.
	
	\bibitem{Pikovsky2003}
	A.~Pikovsky, M.~Rosenblum, and J.~Kurths, ``{Synchronization: A Universal
		Concept in Nonlinear Sciences},'' {\em Cambridge Nonlinear Science Series
		12}, p.~432, 2003.
	
	\bibitem{gray1994}
	C.~M. Gray, ``{Synchronous Oscillations in Neuronal Systems: Mechanisms and
		Functions},'' {\em Journal of Computational Neuroscience}, vol.~1,
	pp.~11--38, 1994.
	
	\bibitem{ottino2016kuramoto}
	B.~Ottino-L{\"o}ffler and S.~H. Strogatz, ``Kuramoto model with uniformly
	spaced frequencies: Finite-n asymptotics of the locking threshold,'' {\em
		Physical Review E}, vol.~93, no.~6, p.~062220, 2016.
	
	\bibitem{Sakaguchi1988}
	H.~Sakaguchi, ``Cooperative phenomena in coupled oscillator systems under
	external fields,'' {\em Progress of Theoretical Physics}, vol.~79, no.~1,
	pp.~39--46, 1988.
	
	\bibitem{Hindes2015}
	J.~Hindes and C.~R. Myers, ``Driven synchronization in random networks of
	oscillators,'' {\em Chaos: An Interdisciplinary Journal of Nonlinear
		Science}, vol.~25, no.~7, p.~073119, 2015.
	
	\bibitem{Lizarraga2020}
	J.~U.~F. Lizarraga and M.~A.~M. de~Aguiar, ``Synchronization and spatial
	patterns in forced swarmalators,'' {\em Chaos (Woodbury, N.Y.)}, vol.~30,
	p.~053112, May 2020.
	
	\bibitem{PhysRevLett.131.030401}
	T.~Murtadho, S.~Vinjanampathy, and J.~Thingna, ``Cooperation and competition in
	synchronous open quantum systems,'' {\em Phys. Rev. Lett.}, vol.~131,
	p.~030401, Jul 2023.
	
	\bibitem{odor2023synchronization}
	G.~{\'O}dor, I.~Papp, S.~Deng, and J.~Kelling, ``Synchronization transitions on
	connectome graphs with external force,'' {\em Frontiers in Physics}, vol.~11,
	p.~1150246, 2023.
	
	\bibitem{tanaka2011multistable}
	T.~Tanaka and T.~Aoyagi, ``Multistable attractors in a network of phase
	oscillators with three-body interactions,'' {\em Physical Review Letters},
	vol.~106, no.~22, p.~224101, 2011.
	
	\bibitem{bick2016chaos}
	C.~Bick, P.~Ashwin, and A.~Rodrigues, ``Chaos in generically coupled phase
	oscillator networks with nonpairwise interactions,'' {\em Chaos: An
		Interdisciplinary Journal of Nonlinear Science}, vol.~26, no.~9, 2016.
	
	\bibitem{battiston2020networks}
	F.~Battiston, G.~Cencetti, I.~Iacopini, V.~Latora, M.~Lucas, A.~Patania, J.-G.
	Young, and G.~Petri, ``Networks beyond pairwise interactions: Structure and
	dynamics,'' {\em Physics Reports}, vol.~874, pp.~1--92, 2020.
	
	\bibitem{dutta2023impact}
	S.~Dutta, A.~Mondal, P.~Kundu, P.~Khanra, P.~Pal, and C.~Hens, ``Impact of
	phase lag on synchronization in frustrated kuramoto model with higher-order
	interactions,'' {\em Physical Review E}, vol.~108, no.~3, p.~034208, 2023.
	
	\bibitem{leon2024higher}
	I.~Le{\'o}n, R.~Muolo, S.~Hata, and H.~Nakao, ``Higher-order interactions
	induce anomalous transitions to synchrony,'' {\em Chaos: An Interdisciplinary
		Journal of Nonlinear Science}, vol.~34, no.~1, 2024.
	
	\bibitem{ganmor2011sparse}
	E.~Ganmor, R.~Segev, and E.~Schneidman, ``Sparse low-order interaction network
	underlies a highly correlated and learnable neural population code,'' {\em
		Proceedings of the National Academy of sciences}, vol.~108, no.~23,
	pp.~9679--9684, 2011.
	
	\bibitem{petri2014homological}
	G.~Petri, P.~Expert, F.~Turkheimer, R.~Carhart-Harris, D.~Nutt, P.~J. Hellyer,
	and F.~Vaccarino, ``Homological scaffolds of brain functional networks,''
	{\em Journal of The Royal Society Interface}, vol.~11, no.~101, p.~20140873,
	2014.
	
	\bibitem{giusti2015clique}
	C.~Giusti, E.~Pastalkova, C.~Curto, and V.~Itskov, ``Clique topology reveals
	intrinsic geometric structure in neural correlations,'' {\em Proceedings of
		the National Academy of Sciences}, vol.~112, no.~44, pp.~13455--13460, 2015.
	
	\bibitem{reimann2017cliques}
	M.~W. Reimann, M.~Nolte, M.~Scolamiero, K.~Turner, R.~Perin, G.~Chindemi,
	P.~D{\l}otko, R.~Levi, K.~Hess, and H.~Markram, ``Cliques of neurons bound
	into cavities provide a missing link between structure and function,'' {\em
		Frontiers in computational neuroscience}, vol.~11, p.~266051, 2017.
	
	\bibitem{sizemore2018cliques}
	A.~E. Sizemore, C.~Giusti, A.~Kahn, J.~M. Vettel, R.~F. Betzel, and D.~S.
	Bassett, ``Cliques and cavities in the human connectome,'' {\em Journal of
		computational neuroscience}, vol.~44, pp.~115--145, 2018.
	
	\bibitem{grilli2017higher}
	J.~Grilli, G.~Barab{\'a}s, M.~J. Michalska-Smith, and S.~Allesina,
	``Higher-order interactions stabilize dynamics in competitive network
	models,'' {\em Nature}, vol.~548, no.~7666, pp.~210--213, 2017.
	
	\bibitem{ghosh2024chimeric}
	R.~Ghosh, U.~K. Verma, S.~Jalan, and M.~D. Shrimali, ``Chimeric states induced
	by higher-order interactions in coupled prey--predator systems,'' {\em Chaos:
		An Interdisciplinary Journal of Nonlinear Science}, vol.~34, no.~6, 2024.
	
	\bibitem{sanchez2019high}
	A.~Sanchez-Gorostiaga, D.~Baji{\'c}, M.~L. Osborne, J.~F. Poyatos, and
	A.~Sanchez, ``High-order interactions distort the functional landscape of
	microbial consortia,'' {\em PLoS Biology}, vol.~17, no.~12, p.~e3000550,
	2019.
	
	\bibitem{benson2016higher}
	A.~R. Benson, D.~F. Gleich, and J.~Leskovec, ``Higher-order organization of
	complex networks,'' {\em Science}, vol.~353, no.~6295, pp.~163--166, 2016.
	
	\bibitem{de2020social}
	G.~F. de~Arruda, G.~Petri, and Y.~Moreno, ``Social contagion models on
	hypergraphs,'' {\em Physical Review Research}, vol.~2, no.~2, p.~023032,
	2020.
	
	\bibitem{iacopini2019simplicial}
	I.~Iacopini, G.~Petri, A.~Barrat, and V.~Latora, ``Simplicial models of social
	contagion,'' {\em Nature communications}, vol.~10, no.~1, p.~2485, 2019.
	
	\bibitem{jhun2019simplicial}
	B.~Jhun, M.~Jo, and B.~Kahng, ``Simplicial sis model in scale-free uniform
	hypergraph,'' {\em Journal of Statistical Mechanics: Theory and Experiment},
	vol.~2019, no.~12, p.~123207, 2019.
	
	\bibitem{vega2004fitness}
	Y.~M. Vega, M.~V{\'a}zquez-Prada, and A.~F. Pacheco, ``Fitness for
	synchronization of network motifs,'' {\em Physica A: Statistical Mechanics
		and its Applications}, vol.~343, pp.~279--287, 2004.
	
	\bibitem{berec2016chimera}
	V.~Berec, ``Chimera state and route to explosive synchronization,'' {\em Chaos,
		Solitons \& Fractals}, vol.~86, pp.~75--81, 2016.
	
	\bibitem{skardal2019abrupt}
	P.~S. Skardal and A.~Arenas, ``Abrupt desynchronization and extensive
	multistability in globally coupled oscillator simplexes,'' {\em Physical
		Review Letters}, vol.~122, no.~24, p.~248301, 2019.
	
	\bibitem{skardal2020higher}
	P.~S. Skardal and A.~Arenas, ``Higher order interactions in complex networks of
	phase oscillators promote abrupt synchronization switching,'' {\em
		Communications Physics}, vol.~3, no.~1, p.~218, 2020.
	
	\bibitem{dai2021d}
	X.~Dai, K.~Kovalenko, M.~Molodyk, Z.~Wang, X.~Li, D.~Musatov, A.~Raigorodskii,
	K.~Alfaro-Bittner, G.~Cooper, G.~Bianconi, {\em et~al.}, ``D-dimensional
	oscillators in simplicial structures: odd and even dimensions display
	different synchronization scenarios,'' {\em Chaos, Solitons \& Fractals},
	vol.~146, p.~110888, 2021.
	
	\bibitem{sarika2024}
	B.~Moyal, P.~Rajwani, S.~Dutta, and S.~Jalan, ``Rotating clusters in
	phase-lagged kuramoto oscillators with higher-order interactions,'' {\em
		Phys. Rev. E}, vol.~109, p.~034211, Mar 2024.
	
	\bibitem{biswas2024symmetry}
	D.~Biswas and S.~Gupta, ``Symmetry-breaking higher-order interactions in
	coupled phase oscillators,'' {\em Chaos, Solitons \& Fractals}, vol.~181,
	p.~114721, 2024.
	
	\bibitem{sayeed2024global}
	M.~Sayeed~Anwar, D.~Ghosh, and T.~Carletti, ``Global synchronization on
	time-varying higher-order structures,'' {\em Journal of Physics: Complexity},
	vol.~5, no.~1, p.~015020, 2024.
	
	\bibitem{muolo2024phase}
	R.~Muolo, T.~Njougouo, L.~V. Gambuzza, T.~Carletti, and M.~Frasca, ``Phase
	chimera states on nonlocal hyperrings,'' {\em Physical Review E}, vol.~109,
	no.~2, p.~L022201, 2024.
	
	\bibitem{ashwin2016hopf}
	P.~Ashwin and A.~Rodrigues, ``Hopf normal form with sn symmetry and reduction
	to systems of nonlinearly coupled phase oscillators,'' {\em Physica D:
		Nonlinear Phenomena}, vol.~325, pp.~14--24, 2016.
	
	\bibitem{leon2019phase}
	I.~Le{\'o}n and D.~Paz{\'o}, ``Phase reduction beyond the first order: The case
	of the mean-field complex ginzburg-landau equation,'' {\em Physical Review
		E}, vol.~100, no.~1, p.~012211, 2019.
	
	\bibitem{fariello2024third}
	R.~Fariello and M.~A. de~Aguiar, ``Third order interactions shift the critical
	coupling in multidimensional {K}uramoto models,'' {\em arXiv preprint
		arXiv:2404.16715}, 2024.
	
	\bibitem{suman2024finite}
	A.~Suman and S.~Jalan, ``Finite-size effect in {K}uramoto oscillators with
	higher-order interactions,'' {\em Chaos: An Interdisciplinary Journal of
		Nonlinear Science}, vol.~34, no.~10, 2024.
	
	\bibitem{Ott2008}
	E.~Ott and T.~M. Antonsen, ``{Low dimensional behavior of large systems of
		globally coupled oscillators},'' {\em Chaos}, vol.~18, no.~3, pp.~1--6, 2008.
	
	\bibitem{kuznetsov1998elements}
	Y.~A. Kuznetsov, {\em Elements of applied bifurcation theory}.
	\newblock Springer, 4~ed., 1998.
	
	\bibitem{VeltzJulia}
	R.~Veltz, ``{BifurcationKit.jl},'' July 2020.
	
\end{thebibliography}

\end{document}